\documentclass[11pt,nofootinbib]{article}
\pdfoutput=1
\usepackage{jheppub}
\usepackage{graphicx}
\usepackage{slashed}
\usepackage[normalem]{ulem}
\usepackage{wasysym}
\usepackage{verbatim}
\usepackage{cancel}
\usepackage{natbib}
\usepackage{amsmath,amssymb,amsfonts}
\usepackage{enumerate}
\usepackage{subfigure} 
\usepackage{caption,subcaption}
\usepackage{tikz}
\usepackage{multirow}
\usetikzlibrary{decorations.pathmorphing, decorations.markings, arrows.meta}
\usepackage{tikz-feynman}
\tikzfeynmanset{compat=1.1.0}

\usepackage{titlesec}
\titleformat*{\subsubsection}{\itshape}

\newcommand\beq{\begin{equation}}
\newcommand\eeq{\end{equation}}

\newcommand\hc{\text{h.c.}}

\def\CP{$CP$}
\def\BR{\text{BR}}
\newcommand{\pt}{p_{\mathrm{T}}}
\newcommand{\dR}{\Delta R}

\newcommand{\PU}[1]{\textbf{\textcolor{blue}{[PU: #1]}}} 

\title{Identifying the origin of the 146-GeV excess at the LHC}
\author[a]{P.\,Uttayarat,}
\author[b]{J.\,Julio}
\author[c]{and R.\,Primulando}

\affiliation[a]{Department of Physics, Srinakharinwirot University, 114 Sukhumvit 23rd Rd., Wattana, Bangkok 10110, Thailand}
\affiliation[b]{National Research and Innovation Agency, Kawasan Sains dan Teknologi B.\,J.\,Habibie, South Tangerang 15314, Indonesia}
\affiliation[c]{Center for Theoretical Physics, Department of Physics, Parahyangan Catholic University, Jl. Ciumbuleuit 94, Bandung 40141, Indonesia}

\emailAdd{patipan@g.swu.ac.th}
\emailAdd{julio@brin.go.id}
\emailAdd{rprimulando@unpar.ac.id}

\abstract{
The 146-GeV excess in the electron-muon final state reported by the CMS Collaboration offers a tantalizing hint for new physics. There are two competing explanations for the origin of the 146-GeV resonance: production via mixing with the Standard Model Higgs boson and leptophilic production through sea electrons and muons inside the proton. In anticipation that such an excess may still persist by the end of LHC Run~3, 
{
we propose three related production channels $H\gamma$, $H j$, and $H jj$, which could be used to distinguish between these two possible scenarios. Taking the two-Higgs-doublet model (2HDM) as an explicit example,
}
we estimate that with $3000~\text{fb}^{-1}$ of the high-luminosity LHC data, the $H\gamma$ channel can provide a striking discriminating power between the mixing and leptophilic scenarios with a significance of 7.1. Combining it with the other two channels, $H j$ and $H jj$, the significance can be pushed up to $7.6$.  
}

\begin{document}
\maketitle

\flushbottom
\section{Introduction}

Following the discovery of the 125-GeV Higgs boson ($h$) last decade, the final missing piece of the Standard Model (SM) was completed. Subsequent measurements have confirmed that the properties of the Higgs boson are consistent with SM predictions~\cite{ATLAS:2016neq,ATLAS:2022vkf,CMS:2022dwd}, further cementing the model's success. However, current experimental data cannot preclude the existence of novel particles or interaction types not prescribed by the SM, leaving room for new physics scenarios. A particularly striking manifestation of this new physics is lepton flavor violation (LFV). Although firmly established in the neutrino sector through oscillations, LFV has not yet been observed among charged leptons. Nevertheless, the lack of experimental signal does not impede interest in searching for these beyond SM interactions at the LHC and other low-energy experiments.

Recently, a hint of LFV at the LHC energy scale has emerged from the CMS search for LFV heavy Higgs decaying to an electron and muon pair~\cite{CMS:2023pte}. This search used 138 fb$^{-1}$ of the LHC Run-2 data in $pp$ collisions at a center-of-mass energy of 13 TeV. The CMS Collaboration performed searches for a new Higgs boson in a range of 110--160 GeV. 
They found a hint of an excess above the background at around $m_{e\mu}=146~\text{GeV}$ with a local (global) significance of 3.8 (2.8) standard deviations.
The excess can be explained by a new particle with a production cross section times $e\mu$ branching ratio of $3.89^{+1.11}_{-1.08}$(stat.)$^{+0.57}_{-0.34}$(syst.) fb. A similar analysis by ATLAS did not report such an excess~\cite{ATLAS:2019old}. However, it must be stressed that ATLAS does not specifically search for a new resonance along the aforementioned $m_{e\mu}$ range. 


The 146-GeV excess in $e\mu$ final states reported by CMS has prompted several studies attempting to explain it in terms of a new resonance. Based on the production mechanisms, the present explanations can be divided into two categories: (a) the scalar mixing scenario~\cite{Primulando:2023ugc,Koivunen:2023led,Gao:2026qnt} and (b) the resonance production for leptophilic (neutral) scalar scenario~\cite{Afik:2023vyl}. 

The mixing scenario, as the name suggests, requires mixing between the 125-GeV Higgs and a new scalar particle so that the latter can be produced via gluon fusion (ggF) and vector-boson fusion (VBF), in the same manner as the SM Higgs boson. To account for the cross section value of $\sigma(pp\to X\to e\mu)=3.89~\text{fb}$, the product of the mixing angle and the corresponding Yukawa couplings is required to be of order $\mathcal{O}(10^{-6})$.  Because of the mixing, direct couplings to quarks, $W^+W^-$, or even same-flavor leptons can be induced. This will lead to other, already searched, LFV processes, most notably the $\mu\to e\gamma$ decay~\cite{Crivellin:2013wna,Crivellin:2014cta}, which could occur at both one- and two-loop levels. 
The UV-completed realization of the mixing scenario in the context of the 2HDM has been discussed in Ref.~\cite{Primulando:2023ugc}. While the $\mu\to e\gamma$ process is induced, its rate can be kept below the current and future projected MEG~II limits~\cite{MEGII:2025gzr} when evaluating the complete set of diagrams~\cite{Altmannshofer:2025nsl}.

The leptophilic scenario, on the other hand, requires that the new scalar have only direct couplings to leptons. Hence, mixing with the 125-GeV scalar is forbidden or extremely small to account for the excess. In the 2HDM realization~\cite{Afik:2023vyl}, the production instead occurs through $e\mu$ annihilation coming from lepton PDF of the proton. Due to the suppression of the lepton PDF, larger LFV couplings are needed, causing them to be particularly sensitive to the LEP dimuon constraint~\cite{OPAL:2003kcu} and muonium-antimuonium oscillation bound~\cite{Willmann:1998gd,Hou:1995dg,Conlin:2020veq,Fukuyama:2021iyw}. 


In this paper, we entertain the possibility that the 146-GeV excess persists after the end of the LHC Run 3. Then, it becomes imperative to determine the origin of the resonance. Thus, we will determine the ability of the high-luminosity (HL) LHC to differentiate between the mixing and the leptophilic production mechanism of the 146-GeV excess. The paper is organized as follows. In Section~\ref{sec:model}, we provide a brief summary of the models for the excess, including their low-energy constraints. We then introduce LHC signatures, which could help distinguish between the mixing and the leptophilic origins of the signal in Section~\ref{sec:distinguish}. Finally, we conclude and discuss our results in Section~\ref{sec:conc}.

\section{Models for 146-GeV excess}
\label{sec:model}
In the present study, we shall provide a way to distinguish the two scenarios in the framework of 2HDM. For convenience, we work in the Higgs basis~\cite{Georgi:1978ri}, where the two Higgs doublets, $H_{1,2}$, are written as
\begin{equation}
	H_1 = \begin{pmatrix}G^+\\[0.2em] \dfrac{v+h_1+iG}{\sqrt{2}} \end{pmatrix},\quad
	H_2 = \begin{pmatrix}H^+ \\[0.2em] \dfrac{h_2+iA}{\sqrt{2}}\end{pmatrix}.
	\label{eq:doublet}
\end{equation}
Here $v=246$ GeV is the electroweak vacuum expectation value, $G^+$ and $G$ are the would-be Goldstone bosons, $h_1$ and $h_2$ are the $CP$-even neutral scalars, $A$ is the pseudoscalar and $H^+$ is the charged Higgs. The dynamics of the scalar sector is governed by the potential 
%
\begin{align}
    V =&~ m_{11}^2(H_1^\dagger H_1) + m_{22}^2(H_2^\dagger H_2) - m_{12}^2(H_1^\dagger H_2 + \hc)  \nonumber \\
    & + \tfrac{1}{2}\lambda_1(H_1^\dagger H_1)^2 + \tfrac{1}{2}\lambda_2(H_2^\dagger H_2)^2 +\lambda_3(H_1^\dagger H_1)(H_2^\dagger H_2) + \lambda_4(H_1^\dagger H_2)(H_2^\dagger H_1) \nonumber \\
    & +\left\{\tfrac{1}{2}\lambda_5(H_1^\dagger H_2)^2 + \left[\lambda_6(H_1^\dagger H_1) + \lambda_7(H_2^\dagger H_2) \right](H_1^\dagger H_2) + \hc \right\}.
    \label{eq:sc-pot}
\end{align}
We shall assume \CP~invariance in the scalar sector, implying that all parameters in Eq.~\eqref{eq:sc-pot} are real. Note that the parameters are not all independent.  The minimization of the potential gives two relations
\begin{align}
    m_{11}^2 = -\tfrac{1}{2}\lambda_1 v^2, \quad m_{12}^2 = \tfrac{1}{2}\lambda_6 v^2,
\end{align}
which can be used to express the physical scalar masses as
\begin{align}
    m_{H^+}^2=&~m_{22}^2 + \tfrac{1}{2}\lambda_3 v^2, \nonumber \\
    m_A^2 =&~ m_{H^+}^2 + \tfrac{1}{2}(\lambda_4-\lambda_5)v^2.
\end{align} 
The two \CP-even states, $h_1$ and $h_2$, mix via
\begin{equation}
	\begin{pmatrix}h_1\\h_2\end{pmatrix} = \begin{pmatrix}c_\alpha &s_\alpha\\-s_\alpha &c_\alpha\end{pmatrix} \begin{pmatrix}h\\H\end{pmatrix}, 
 \label{eq:mix}
\end{equation}
where $h$ is the 125-GeV Higgs boson, $H$ is the heavier scalar, and $c_\alpha$ ($s_\alpha$) stands for $\cos \alpha$ ($\sin \alpha$). The mixing is induced by the quartic coupling $\lambda_6$ as
\begin{align}
    s_{2\alpha} = \frac{2\lambda_6 v^2}{m_H^2-m_h^2}.
\end{align}
In the limit of $s_\alpha\ll 1$, we can approximate
\begin{align}
    m_h^2 =&~ \lambda_1 v^2 + \mathcal{O}(s_\alpha^2), \nonumber \\
    m_H^2 =&~ m_{H^+}^2 + \tfrac{1}{2}(\lambda_4+\lambda_5)v^2 + \mathcal{O}(s_\alpha^2).
\end{align}

The presence of $m_{22}^2$ implies that the $H,A,H^+$ masses can be well above the electroweak scale. However, their mass differences are restricted by the electroweak precision measurements, particularly the $T$ parameter. In the limit of small $s_\alpha$, the shift $\Delta T$ induced by new physics contributions is given by~\cite{Grimus:2008nb,Primulando:2022vip}
\begin{align}
    \Delta T = \frac{1}{16\pi^2\alpha_{em}v^2}\left[F(m_H^2,m_{H^+}^2)+F(m_A^2,m_{H^+}^2)-F(m_H^2,m_A^2)\right],
    \label{eq:T-par}
\end{align}
where $\alpha_{em}=1/137$ denotes the electromagnetic fine-structure constant in the Thomson limit and
\begin{align}
    F(x,y) = \frac{x+y}{2} - \frac{xy}{x-y}\ln\frac{x}{y}.
    \label{eq:T-func}
\end{align}
The correction $\Delta T$ vanishes if one of the neutral scalars is degenerate with the charged scalar, a condition consistent with the experimental value of $\Delta T=0.00\pm0.06$~\cite{ParticleDataGroup:2024cfk}. The scalar mass differences are further restricted by theoretical constraints on the quartic couplings $\lambda_i$. For the model to be valid, these couplings must be perturbative, as well as preserve vacuum stability~\cite{Ivanov:2006yq} and partial wave unitarity~\cite{Ginzburg:2003fe}. 


The Yukawa interactions of $H_1$ give rise to fermion masses, while those of $H_2$ generate potential flavor-violating interactions. It can be assumed that $H_2$ does not have direct couplings with quarks, so interactions with leptons can be written explicitly as
\begin{align}
    \mathcal{L} \supset & -\frac{\sqrt{2}m_i}{v}\bar{L}_ie_{Ri}H_1 - \sqrt{2}Y_{ij}\bar{L}_ie_{Rj}H_2 + \hc,
\end{align}
where $L$ and $e_R$ are the usual lepton doublet and singlet, respectively, $m_i$ is the lepton mass of the $i$-th generation, and $i,j=e,\mu,\tau$ denote the lepton generations. The two scenarios that we are about to discuss require  nonvanishing $Y_{e\mu}$ and/or $Y_{\mu e}$ to generate the process in question. Generally, these two couplings can be complex, but thanks to the strong constraint from electric dipole moment of the electron, the two phases must be practically aligned. For simplicity, we set both phases to zero. Other $Y_{ij}$ couplings, although not directly involved in the production process, may be present.

In this work, we associate the CMS excess with the production of $H$ scalar, hence $m_H=146~\text{GeV}$.  Let us suppose for the moment that only $Y_{e\mu}$ and $Y_{\mu e}$ are nonzero, so that the charged scalar $H^\pm$ will decay exclusively into an electron/muon and missing energy, which places a lower bound of $550~\text{GeV}$ on its mass~\cite{CMS:2018eqb,ATLAS:2019lff}. To satisfy the electroweak precision data, $m_A$ must lie close to $m_{H^+}$. Therefore, in this study, we take $m_A=m_{H^+}$ to ensure $\Delta T=0$. 

The two neutral scalars $H$ and $A$ can be pair produced via the Drell-Yan process. Their mass hierarchy allows the decay channel $A\to HZ$ to open, alongside $A\to e\mu$. Meanwhile, $H$ decays dominantly into $e\mu$ for a sufficiently small mixing angle  (i.e., $s_\alpha\lesssim0.025$). As a result, the pair-produced $H$ and $A$ give rise to four-lepton (or more) final states, which are severely constrained by the CMS multilepton search~\cite{CMS:2021cox}. In this scenario, the analyses of Refs.~\cite{Primulando:2023ugc,Uttayarat:2025qjl} place a strong constraint on the mass of $A$, with $m_A\gtrsim 760~\text{GeV}$. Such a heavy $A$ will lead to severe tension with perturbativity and vacuum stability constraints of the scalar potential. To relax the lower bound on $m_A$, one needs to dilute the $H\to e\mu$ branching ratio, either by choosing a larger $s_\alpha$, so that other $H$ decay channels (e.g., $WW^*$, $ZZ^*$, and $b\bar{b}$) become increasingly important, or by introducing a less-visible $H$ decay mode (such as $H\to \tau^+\tau^-$). 

Finally, the direct upper bound on $s_\alpha$ is determined from the single $H$ production searches at CMS for a heavy resonance decaying into $ZZ^\ast$~\cite{CMS:2018amk} and $\tau^+\tau^-$~\cite{CMS:2022goy}. Both channels constrain the production cross section to less than $\mathcal{O}(100)~\text{fb}$, corresponding to  $s_\alpha\lesssim 0.2$. This bound is consistent with the combined ATLAS and CMS Run 1 and Run 2 Higgs data fits, which yield $s_\alpha \leq0.15$~\cite{ATLAS:2016neq,ATLAS:2022vkf,CMS:2022dwd}.

\subsection{Mixing scenario}
In this interpretation, the 146-GeV excess is due to the production of the $H$ resonance via ggF and VBF. Assuming that only $Y_{e\mu}$ and $Y_{\mu e}$ are nonzero, the product of the production cross section and branching ratio is given by~\cite{Primulando:2023ugc}
\begin{align}
    \sigma \times \BR(H\to e\mu) = \frac{s_\alpha^2 \sigma_{SM}\Gamma(H \to e\mu)}{s_\alpha^2\Gamma_{SM} + \Gamma(H\to e\mu)},
    \label{eq:mix-prod}
\end{align}
where $\sigma_{SM}$ and $\Gamma_{SM}$ are the would-be SM Higgs production cross-section and the total decay width for $m_H=146~\text{GeV}$, respectively, while $\Gamma(H\to e\mu)$ is given as
\begin{align}
    \Gamma(H\to e\mu) = \frac{c_\alpha^2 m_H}{8\pi}(Y_{e\mu}^2 + Y_{\mu e}^2). 
\end{align}

The combination of LFV couplings $(Y_{\mu e},Y_{e\mu})$ and $s_\alpha$ will induce the $\mu\to e\gamma$ process, with a decay rate given by
\begin{align}
    \Gamma(\mu \to e\gamma) = \frac{\alpha_{em} m_\mu^5}{256\pi^4 v^4}\left(|c_L|^2 + |c_R|^2\right).
    \label{eq:mu-e-gamma}
\end{align}
The Wilson coefficients $c_{L,R}$ are determined by evaluating the full one- and two-loop diagrams. These include the usual one-loop penguin diagrams, two-loop Barr-Zee diagrams, and those induced by kinetic terms and quartic couplings. The full expressions of these contributions are given in Ref.~\cite{Altmannshofer:2025nsl}. 

The diagrams induced by quartic couplings deserve a detailed discussion due to their significant impact on the decay rate. They are induced by $\lambda_3$ and $\lambda_7$ and appear at the two-loop level, analogous to the top-quark Barr-Zee diagrams, but with the top quark lines replaced by the charged Higgs boson ($H^\pm$) lines. The Wilson coefficients with the photon exchange are given by
\begin{align}
    c_{L}^{H^\pm\gamma} =&~ \frac{\alpha_{em}}{4\pi}\frac{v}{m_\mu}Y_{e\mu}v^2\left[c_\alpha s_\alpha\lambda_3\left(\frac{f_9(x_{H^\pm H})}{m_H^2} - \frac{f_9(x_{H^\pm h})}{m_h^2} \right)\right.\nonumber\\
    &\hspace{3.cm}\left.+ \lambda_7\left(c_\alpha^2\frac{f_9(x_{H^\pm H})}{m_H^2} + s_\alpha^2\frac{f_9(x_{H^\pm h})}{m_h^2}\right)\right],  \\
    c_R^{H^\pm\gamma} =&~ c_L^{H^\pm\gamma}\big|_{Y_{e\mu}\to Y_{\mu e}},
\end{align}
where the loop function $f_9$ is given in Ref.~\cite{Altmannshofer:2025nsl} and $x_{ab} = m_a^2/m_b^2$.
Other contributions from $Z$ and $W$ exchanges also share a similar structure. Notice that terms proportional to $\lambda_3$ are always suppressed by the mixing angle, appearing as the destructive interference between the $h$ and $H$ loops, whereas those proportional to $\lambda_7$ are not. Consequently, a nonzero $\lambda_7$ could counter the dominant Barr-Zee contributions, making its value central to keeping the $\mu\to e\gamma$ decay rate under control. 

As previously noted, a relatively small mixing angle strictly constrains $m_A$. Consistency with vacuum stability requires a lower $m_A$ value. For example, $m_A=m_{H^+}=600~\text{GeV}$ is permissible if we allow other decay modes to thrive by taking a larger $s_\alpha\gtrsim0.025$~\cite{Primulando:2023ugc}. 

\begin{figure}
    \centering
    \includegraphics[width=12cm]{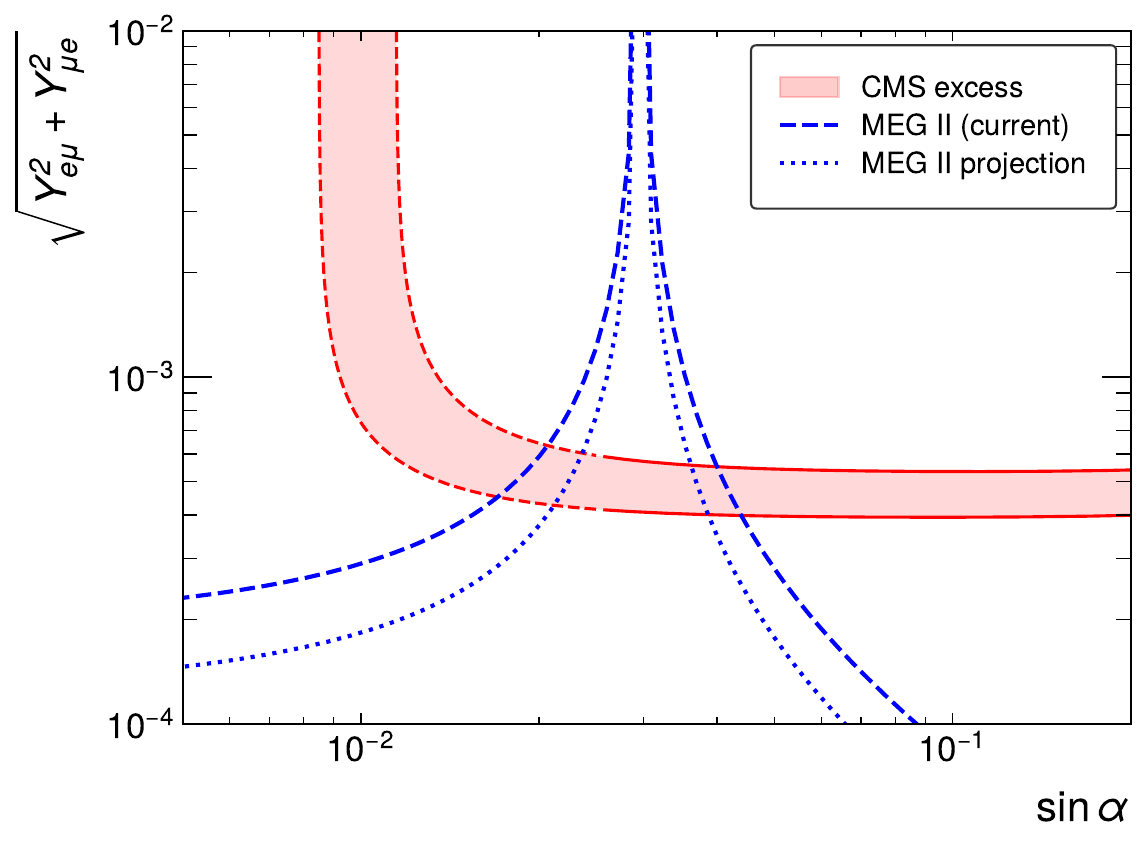}
    \caption{The allowed region of the $146$-GeV CMS excess (red) at the $1\sigma$ level. The current (future) $\mu\to e\gamma$ bound from the MEG II experiment is shown as a dashed (dotted) blue line. The area bounded by the dashed red lines is excluded by the CMS multilepton constraints, unless $m_A\gtrsim760~\text{GeV}$. }
    \label{fig:CMS}
\end{figure}

We present the allowed region consistent with the CMS excess and the $\mu\to e\gamma$ bound in Fig.~\ref{fig:CMS}. In addition to the aforementioned masses, we set $\lambda_3=10$ (as dictated by vacuum stability for our choice of scalar masses) and $\lambda_7=-0.2$. The presence of $\lambda_7$ is essential to evade the $\mu\to e\gamma$ constraint, as indicated by the blind spot around $s_\alpha=0.03$. The relatively small value of $\lambda_7$ allows the vacuum-stability constraints to be satisfied.

\subsection{Leptophilic scenario}
This scenario assumes that $H$ does not couple to quarks at all, implying a vanishing mixing angle $\alpha$. As explained in Ref.~\cite{Afik:2023vyl}, the production occurs via $e\mu$ collision coming from the lepton PDFs of the proton, and thus requires $Y_{e\mu},Y_{\mu e}\sim \mathcal{O}(1)$. The only direct bounds on these couplings come from LEP dimuon and the muonium-antimuonium constraints, restricting $Y_{e\mu},Y_{\mu e}\lesssim0.7$. 

Owing to the zero mixing, the region with lighter $m_A$ also suffers from strong CMS multilepton constraints. To loosen the bound on $m_A$, a new decay channel, i.e., $H\to\tau^+\tau^-$, is needed, which in turn induces $\mu\to e\gamma$ at the two-loop level. Similar to the mixing case, this LFV bound can be evaded by including $\lambda_7$. For instance, the reported CMS excess and the current MEG II data can be satisfied for $m_A=m_{H^+}=600~\text{GeV}$ by having $Y_{e\mu}=0.7/\sqrt{2}$, $Y_{\tau\tau}=-0.4/\sqrt{2}$, $\lambda_3=10$, and $\lambda_7=5.2$. A consistency check with vacuum stability and perturbativity requires a more involved analysis, which is beyond the scope of this paper. 


\section{Distinguishing among the possible origins of the 146-GeV excess}
\label{sec:distinguish}
In this section, we discuss additional collider signatures that can help distinguish between the two explanations of the 146-GeV excess: associated production with a photon ($H\gamma$), associated production with a jet ($H j$), and VBF production ($H jj$). In each case, we simulate signal and background events at $\sqrt{s} = 14$~TeV using Madgraph5~\cite{Alwall:2014hca}. 
{For leptophilic signal events that have leptons in the initial states, we use the LUXlep-NNPDF31 parton distribution function (PDF)~\cite{Buonocore:2020nai}. For other signal and background events, we use the NN23LO1 PDF set~\cite{Ball:2012cx}. The parameter space of the signals is normalized such that the cross section times branching fraction is $3.89$ fb at $\sqrt{s} = 13$ TeV. The events are then passed to Pythia8~\cite{Bierlich:2022pfr} for parton showering and hadronization. }
Detector simulation and minimum-bias modeling (with an average pileup of 200) are handled by \textsc{Delphes}~3~\cite{deFavereau:2013fsa}. UFO model files for both signal cases are generated with Feynrules~2.3~\cite{Alloul:2013bka}. 


In our simulations, we require an event to contain exactly one oppositely charged electron-muon pair. 
Specifically, both leptons must have $p_T > 25$ GeV. 
The pseudorapidity of the electron (muon) must satisfy $|\eta| < 2.47$ (2.4). 
Electron and muon candidates are required to be isolated from other
activity in the event. For each candidate, the isolation variable is
defined as the scalar sum of the transverse momenta of all reconstructed
particles with $\pt > 0.5$ GeV within a cone of size
$\dR = 0.5$ around the candidate direction, excluding the candidate itself, divided by the transverse momentum of the candidate. The relative isolation is required to be less than $0.12$ for electrons and less than $0.25$ for muons. Furthermore, to suppress the backgrounds involving top quarks, we reject any events containing a $b$-tagged 
jet with $p_T > 30$ GeV within the tracker acceptance $|\eta| < 2.5$.%


{Since our goal is to distinguish between the mixing and the leptophilic  scenarios, we define $\mu_1 = N_\text{mixing} + N_\text{SM}$ and $\mu_0 = N_\text{leptophilic} + N_\text{SM}$. We use the $N_{\text{leptophilic}} + N_{\text{SM}}$ as the null hypothesis since this scenario yields a lower, more conservative significance. We quantify the separation between the two scenarios with the Poisson likelihood-ratio test statistic, in which the significance for discriminating between the two hypotheses in a given signal region is~\cite{Cowan:2010js}
\begin{equation} \label{eq:sigmadef}
  Z = \sqrt{\,2\left[\mu_1\ln\!\left(\frac{\mu_1}{\mu_0}\right)
      - \left(\mu_1-\mu_0\right)\right]}\,.
\end{equation}
}


\subsection{Associated production with photon ($H\gamma$)}
\begin{figure}[t!]
    \centering
    
    \subfigure[\label{fig:feynman}]{
        \begin{tikzpicture}[
            thick, 
            scale=0.8, every node/.style={transform shape},
            gluon/.style={decorate, draw=black,
                decoration={coil, amplitude=4pt, segment length=5pt}},
            scalar/.style={dashed, draw=black},
            fermion/.style={draw=black, postaction={decorate},
                decoration={markings, mark=at position 0.6 with {\arrow{Stealth}}}},
            antifermion/.style={draw=black, postaction={decorate},
                decoration={markings, mark=at position 0.6 with {\arrowreversed{Stealth}}}},
            photon/.style={decorate, draw=black,
                decoration={snake, amplitude=2.5pt, segment length=5pt}},
            vertex/.style={draw, fill=black, circle, inner sep=2pt},
        ]
            \coordinate (g1) at (-2, 1.5);
            \coordinate (g2) at (-2, -1.5);
            \coordinate (v1) at (0, 0);
            \coordinate (v2) at (2, 0);
            \coordinate (v3) at (3, 1);
            \coordinate (f1) at (4, 2);
            \coordinate (gamma) at (4.5, 0.8);
            \coordinate (f2) at (3.5, -1.5);
            \draw[gluon] (g1) node[left] {$g$} -- (v1);
            \draw[gluon] (g2) node[left] {$g$} -- (v1);
            \draw[scalar] (v1) -- (v2) node[midway, above] {$H$};
            \draw[fermion] (v2) -- (v3);
            \draw[fermion] (v3) -- (f1) node[right] {$e^-/\mu^-$};
            \draw[photon] (v3) -- (gamma) node[right] {$\gamma$};
            \draw[antifermion] (v2) -- (f2) node[right] {$\mu^+/e^+$};
            \node[vertex] at (v1) {};
        \end{tikzpicture}
    }
    %
    %
    \hfill
    \subfigure[\label{fig:memu}]{
        \begin{tikzpicture}[
            thick, 
            scale=0.8, every node/.style={transform shape},
            gluon/.style={decorate, draw=black,
                decoration={coil, amplitude=4pt, segment length=5pt}},
            scalar/.style={dashed, draw=black},
            fermion/.style={draw=black, postaction={decorate},
                decoration={markings, mark=at position 0.6 with {\arrow{Stealth}}}},
            antifermion/.style={draw=black, postaction={decorate},
                decoration={markings, mark=at position 0.6 with {\arrowreversed{Stealth}}}},
            photon/.style={decorate, draw=black,
                decoration={snake, amplitude=2.5pt, segment length=5pt}}
        ]
            \coordinate (g1) at (-2, 2);
            \coordinate (g2) at (-2, -1.5);
            \coordinate (v1) at (0, 0);
            \coordinate (v2) at (2, 0);
            \coordinate (v3) at (-1, 1);
            \coordinate (f1) at (4, 1.5);
            \coordinate (gamma) at (0.5, 1.7);
            \coordinate (f2) at (3.5, -1.5);
            \draw[fermion] (g1) node[left] {$e^-/\mu^-$} -- (v3);
            \draw[fermion] (v3) -- (v1);
            \draw[fermion] (v1) -- (g2) node[left] {$\mu^+/e^+$};
            \draw[photon] (v3) -- (gamma) node[right] {$\gamma$};
            \draw[scalar] (v1) -- (v2) node[midway, above] {$H$};
            \draw[fermion] (v2) -- (f1) node[right] {$e^-/\mu^-$};
            \draw[antifermion] (v2) -- (f2) node[right] {$\mu^+/e^+$};
        \end{tikzpicture}
    }
    
    \caption{Feynman diagrams for the associated production with photon process. (a) The ggF channel where the photon is radiated off the final-state lepton. (b) The $e\mu$ production channel where the photon can also be radiated off from initial-state lepton. 
    }
    \label{fig:feyndiag}
\end{figure}
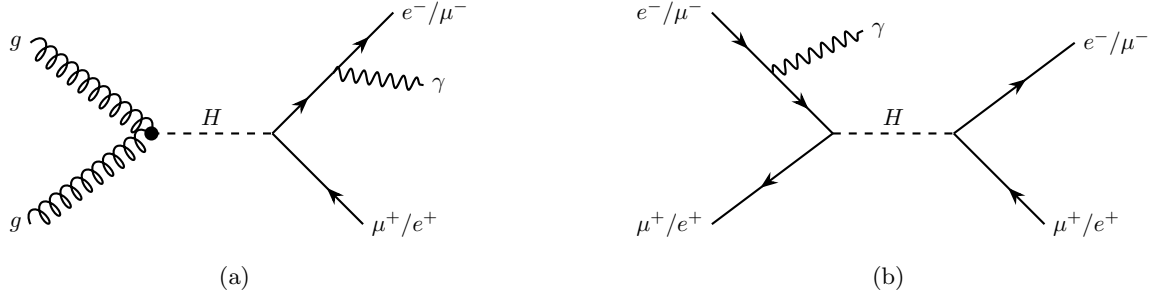

The associated production of a photon with the $e\mu$ pair can provide a way to distinguish between the two explanations for the 146-GeV excess. In the mixing scenario, where the dominant production mode is ggF, the photon must be radiated off from the final-state lepton (FSR). On the other hand, in the leptophilic case, the photon can also be radiated off the initial-state lepton (ISR). The Feynman diagrams for both scenarios are shown in Fig.~\ref{fig:feyndiag}. 

In our analysis, we define an event as a final state containing exactly an $e\mu$ pair and a photon. Furthermore, the photon is required to satisfy  $p_T > 20$ GeV with $|\eta| < 2.37$, excluding the transition region $1.37 < |\eta| < 1.50$. Photon candidates are required to satisfy an isolation requirement
analogous to that applied to the charged leptons. The pile-up-corrected scalar sum of the transverse momenta of all particles with $\pt > 0.5$ GeV within a
cone of size $\dR = 0.5$ around the photon direction, excluding the photon itself, is required to be less than $12\%$ of the photon transverse momentum.

We define two signal regions:
\begin{itemize}
    \item \textbf{SR~A}: $\left|m_{e\mu} - 146\text{ GeV}\right| < 2.5$ 
    GeV, targeting events in which the photon mostly comes from the ISR. In this case, the $e\mu$ invariant mass reconstructs the Higgs mass.
    \item \textbf{SR~B}: $\left|m_{e\mu\gamma} - 146\text{ GeV}\right| < 
    2.5$ GeV, targeting events in which the photon comes from the FSR so that the full $e\mu\gamma$ system reconstructs the 
    Higgs mass. In this signal region, we also require the transverse momentum of the $e\mu\gamma$ 
    system to satisfy $p_T(e\mu\gamma) < 10$ GeV. This cut on 
$p_T(e\mu\gamma)$ is used to distinguish between the leptophilic and mixing scenarios. In the mixing scenario, the system can easily recoil against an 
    ISR jet.
\end{itemize}
The two signal regions are mutually complementary, signal events predominantly populate one or the other region depending on the photon kinematics. Additionally, we veto events containing jets that satisfies $Hj$ and $Hjj$ signal regions described in the following subsections. 


\begin{figure}[t!]
     \centering
         \includegraphics[width=0.45\textwidth]{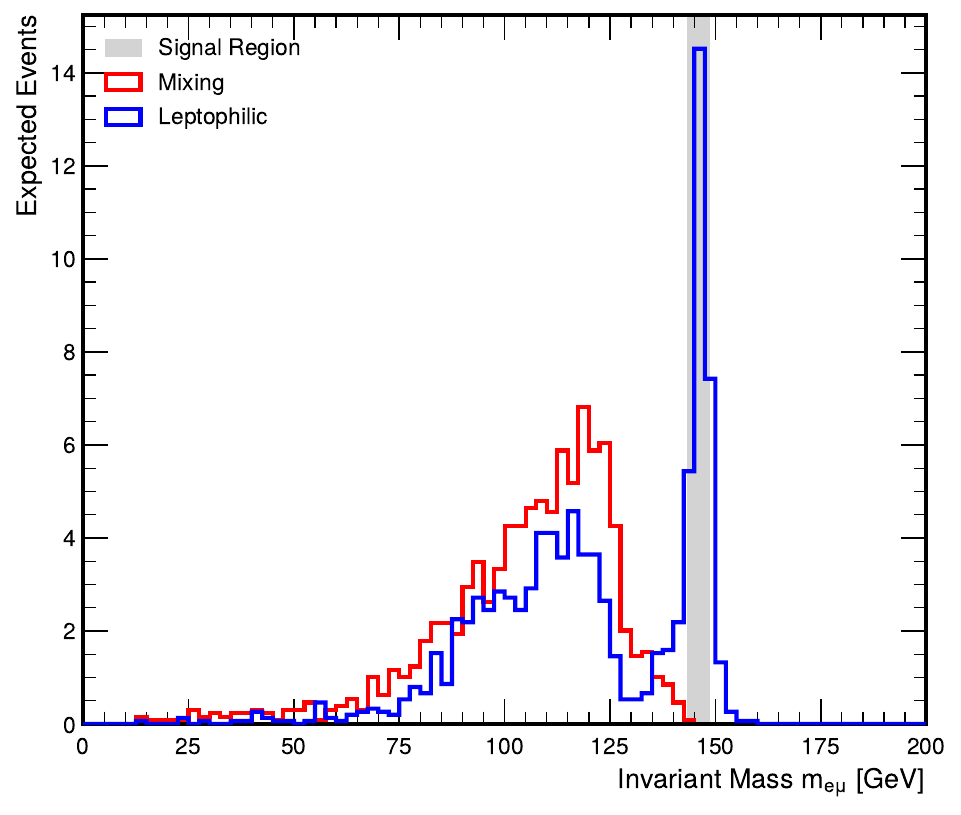}
         \includegraphics[width=0.45\textwidth]{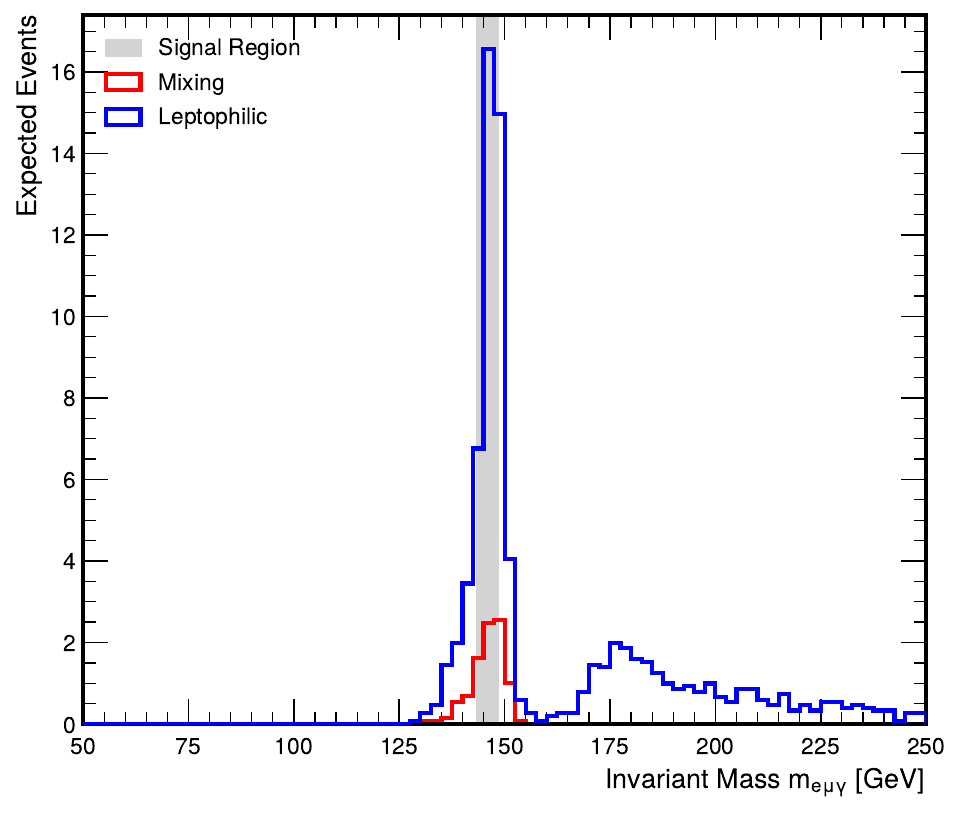}
        \caption{Simulated kinematic distributions of $m_{e\mu}$ (left) and 
        $m_{e\mu\gamma}$ (right) after applying all kinematic selections, except the invariant mass, for the mixing and leptophilic scenarios at $\sqrt{s} = 14$ TeV with 
        the average pile up of 200. The gray band in both figures denotes the signal region. In the right panel, the mixing scenario yields lower values since the cut $p_T(e\mu\gamma) < 10$ GeV has been imposed.}
        \label{fig:massdist}
\end{figure}

The simulated invariant mass distributions after the baseline selection are shown in Fig.~\ref{fig:massdist}. A clear peak near $m_{e\mu} \approx 146$ GeV is visible in the SR~A, while for SR~B the corresponding peak is around $m_{e\mu\gamma}\approx 146$ GeV. The simulated backgrounds for both signal regions are $W^+W^-\gamma$ and $t \bar t \gamma$, in which the $W$-pair decays to different lepton flavors. The expected event yields at the HL-LHC with an integrated luminosity of $\mathcal{L} = 3000\ \text{fb}^{-1}$ are summarized in Table~\ref{tab:exp}. 

\begin{table}[t]
    \centering
    \caption{Expected signal and SM background yields in $H\gamma$, $H j$, and $H jj$  signal regions 
    at $\mathcal{L} = 3000\ \text{fb}^{-1}$.}
    \begin{tabular}{|c|l|c|c|c|}
    \hline
         {\bf Channel} & {\bf Scenario} & \multicolumn{2}{c|}{\bf Yield} & $Z$ \\ \hline
         \multirow{4}{*}{$H\gamma$} &  & {SR~A} & SR~B & \multirow{4}{*}{$7.1$}\\ \cline{3-4} 
          & Mixing & $0.0$ & $4.5$ &  \\ 
         & Leptophilic & $22.0$ & $27.6$  &  \\
         & SM & $3.9$ & $7.2$ & \\ \hline\hline  
         \multirow{3}{*}{$H j$} & Mixing & \multicolumn{2}{c|}{$30.2$} & \multirow{3}{*}{$0.6$} \\
         & Leptophilic & \multicolumn{2}{c|}{$10.1$} & \\
         & SM & \multicolumn{2}{c|}{$1068.8$} & \\ \hline\hline 
         \multirow{3}{*}{$H jj$} & Mixing & \multicolumn{2}{c|}{$186.5$} & \multirow{3}{*}{$2.8$} \\
         & Leptophilic & \multicolumn{2}{c|}{$148.2$} & \\
         & SM & \multicolumn{2}{c|}{$21.2$} & \\ \hline \hline
         \multicolumn{4}{|c|}{\bf Total} & 7.6 \\ \hline
    \end{tabular}
    \label{tab:exp}
\end{table}


We use Eq.~\eqref{eq:sigmadef} to determine the significance of the difference between the mixing and the leptophilic scenarios for 
each signal region independently to get
$Z_{\text{A}} = 5.4$ and $Z_{\text{B}} = 4.5$. The combined significance is found via 
\begin{equation}
    Z_{\text{comb}} = \sqrt{Z_{\text{A}}^2 + 
    Z_{\text{B}}^2} \approx 7.1.
    \label{eq:sigma}
\end{equation}
This result demonstrates that the $\gamma + H$ channel alone, combining the two signal regions, can provide sufficient sensitivity to distinguish between the mixing and leptophilic origins of the 146-GeV excess at the HL-LHC. 

\subsection{Associated production with jet ($H j$)}
\label{sec:emuj}
The $Hj$ channel offers another avenue for discriminating between the mixing 
and leptophilic production scenarios. In the mixing case, the Higgs boson is 
predominantly produced via gluon fusion, which naturally yields events with a 
hard initial-state radiation jet recoiling against the $e\mu$ system. In 
contrast, the leptophilic production mode, proceeding through $e\mu \to H$, 
produces comparatively fewer high-$p_T$ jets, which predominantly arise from the minimum bias events. We exploit this difference by 
requiring at least one energetic jet in addition to the opposite-sign 
electron-muon pair.


Events are required to satisfy the following basic kinematic cuts. The invariant mass of the $e\mu$ system is required to lie within  $\left|m_{e\mu} - 146\text{ GeV}\right| < 2.5$ GeV. We further require  the event to contain at least one jet with $p_T > 110$ GeV and  $|\eta| < 2.5$. We veto events that contain additional jets with $p_T > 50$ GeV and $|\eta| < 5.0$, so that this signal region is mutually exclusive with the VBF-inspired signal region described in the next subsection. 


The expected event yields at $\mathcal{L} = 3000\ \text{fb}^{-1}$ are 
shown in Table~\ref{tab:exp}. The background arises 
from $t\bar{t}$ and $t\bar t j$ production channels. Additionally, $W^+ W^-j$ 
 contributes to the background.

As in the associated production with photon channel, we quantify the ability to distinguish the 
two scenarios for the 146-GeV excess using Eq.~\eqref{eq:sigmadef}. We find $Z \approx 0.6$ at $3000\ \text{fb}^{-1}$. While the monojet 
channel alone does not reach discovery-level sensitivity, it provides 
complementary discriminating power that can be combined with the 
photon associated production result. The sensitivity in this channel is primarily limited by 
the large $t\bar{t}$ background, which dominates the event yield even after 
the $b$-jet veto.

\subsection{VBF-inspired channel ($H jj$)}
The VBF-inspired channel provides a particularly clean environment to 
distinguish between the mixing and leptophilic scenarios. In the mixing case, the 
Higgs boson can be produced through VBF, yielding the characteristic 
topology of two forward jets with a large pseudorapidity gap and high dijet 
invariant mass. The leptophilic production mode, on the other hand, does not 
proceed through VBF and enters the signal region only when accompanied by 
minimum bias events that accidentally mimic the VBF topology. We exploit this 
topological difference by imposing standard VBF selection criteria on the 
dijet system.


The VBF topology is selected by requiring at least two jets with $p_T > 50$ 
GeV and $|\eta| < 5.0$. Among these, the two highest-$p_T$ jets must satisfy 
the following criteria: the leading jet must have $p_T > 80$ GeV, the two 
jets must lie in opposite hemispheres ($\eta_{j_1} \cdot \eta_{j_2} < 0$), 
their pseudorapidity separation must satisfy $|\Delta\eta_{jj}| > 4.0$, and 
the dijet invariant mass must exceed $m_{jj} > 850$ GeV.


The expected event yields at the HL-LHC with $\mathcal{L} = 3000\ \text{fb}^{-1}$ are shown 
in Table~\ref{tab:exp}. The SM background for this channel is $t \bar t (+jj)$ and $W^+W^-jj$. These backgrounds are 
substantially reduced by the stringent dijet requirements and $b$-jet veto.

By using the same figure of merit as in the previous channels, the discriminating power is found as $Z = 2.8$. 
The VBF-inspired channel 
thus provides stronger discriminating power than the monojet channel, owing 
to the much lower background contamination after the dijet topology cuts. 
However, we note that the leptophilic yield remains sizable in this signal 
region, as minimum bias events can produce forward jets that pass the VBF 
selection. We expect that improved pile-up mitigation at the HL-LHC is expected to reduce the leptophilic yield in this signal region beyond what is captured by our simplified Delphes simulation. 

{Combining all the channels, we find a total significance of $Z = 7.6$, which would be sufficient to distinguish the mixing and leptophilic scenarios. This number accounts for statistical uncertainties only. A full evaluation of the experimental systematic uncertainties is beyond the scope of this work, but we can estimate their impact by allowing the expected yield of the null hypothesis in each signal region to float within a Gaussian constraint and profiling it out of the likelihood ratio~\cite{Canonero:2026pul}. Assigning a flat $20\%$ uncertainty to the expected yields, taken as uncorrelated between signal regions, the combined significance is reduced to $5.4$.} 


\section{Conclusion and discussion}
\label{sec:conc}


The nature of the 146-GeV excess in the $e\mu$ final state remains a mystery. If the signal continues to persist after the LHC Run-3 data, then it is imperative to identify its origin. In the literature, there are two proposed production mechanisms for the 146-GeV resonance: the mixing scenario, where the resonance is produced via mixing with the SM Higgs boson, and the leptophilic scenario, where it is produced directly from the sea electrons and muons inside the proton. To differentiate between these two cases, we have proposed three new search channels: $H\gamma$, $H j$, and $H jj$. These three channels target the differences in the initial states of the two scenarios. In the mixing case, the main production modes are ggF and VBF, which are hadronic.  In the leptophilic case, as the name suggests, the initial state is leptonic. We then investigate the ability of these three signatures to distinguish between the mixing and leptophilic origins of the 146-GeV excess at the HL-LHC with an integrated luminosity of $3000~\text{fb}^{-1}$. 

For the $H\gamma$ channel, we introduce two signal regions: SR~A and SR~B. SR~A targets the photon originating from ISR by requiring $|m_{e\mu}-146\text{ GeV}|\le 2.5$ GeV. SR~B targets the FSR photon by requiring $|m_{e\mu\gamma}-146\text{ GeV}|\le 2.5$ GeV. In addition to the cut on $m_{e\mu\gamma}$, we also make a cut on the $p_T$ of the $e\mu\gamma$ system, i.e., $p_T(e\mu\gamma)<10$ GeV. This cut helps suppress signal events in the mixing scenario, where the system tends to recoil against ISR jets. Hence, both signal regions are expected to contain a sizable leptophilic contribution, while one expects minimal events in both SRs if the excess is due to mixing with the Higgs boson. All the cuts mentioned above also help suppress the background events, and thereby enhancing the separation between the two scenarios. If the excess prevails, our analysis indicates that this channel at the HL-LHC with $3000~\text{fb}^{-1}$ integrated luminosity can definitively distinguish between the two production scenarios with a remarkable significance of $7.1$.

In the $H j$ channel, the jet arises from initial-state radiation off the hadronic initial state. Thus, one expects this channel to be sensitive to the mixing scenario. However, due to pileups, there could be an accidental jet in the leptophilic case as well. As in the $e\mu\gamma$ scenario, we make a cut on the invariant mass of the $e\mu$ system, $|m_{e\mu}-146\text{ GeV}|\le 2.5$ GeV. We also require that the jet is hard, with $p_T>110$ GeV. Based on our projection, this channel, dominated by the SM backgrounds, gives a relatively low discriminating power with $Z=0.6$.

The $H jj$ channel is also sensitive to the mixing scenario. In addition to the cut on the $e\mu$ invariant mass, $|m_{e\mu}-146\text{ GeV}|\le 2.5$ GeV, we also require two well-separated jets with $p_T>50$ GeV and $m_{jj}>850$ GeV. The requirement on the dijet system helps suppress the SM background. We estimate that the $H jj$ search channel can distinguish between the leptophilic and mixing scenarios with a discriminating power of $Z=2.8$. 

In conclusion, we find that the two possible explanations for the CMS 146-GeV excess can be robustly distinguished at the HL-LHC with the integrated luminosity of $3000~\text{fb}^{-1}$. By combining the statistical significance across all three channels, we obtain a total discriminating power of $Z=7.6$. Even with a 20\% systematic uncertainty, it is expected that the HL-LHC can still discriminate between the two scenarios at a significance of around $5\sigma$. This level of sensitivity demonstrates that the upcoming run of the HL-LHC has great potential to conclusively unravel the true nature of this excess.

\acknowledgments{ 
R.P. was supported by Direktorat Penelitian dan Pengabdian kepada Masyarakat, Direktorat Jenderal Riset dan Pengembangan, Kementerian Pendidikan Tinggi, Sains dan Teknologi
Republik Indonesia in the year 2025 with contract number 7939/LL4/PG/2025; III/LPPM/ 2025-06/154-PE and 125/C3/DT.05.00/PL/2025. The work of P.U. was supported in part by Thailand NSRF via PMU-B under grant number B39G690007. R.P. and P.U. also acknowledged the National Science and Technology Development Agency, National e-Science Infrastructure Consortium, Chulalongkorn University and the Chulalongkorn Academic Advancement into Its 2nd Century Project, NSRF via the Program Management Unit for Human Resources \& Institutional Development, Research and Innovation (Thailand) [grant numbers B39G680009] for providing computing infrastructure that has contributed to the research results reported within this paper.
}

\

\bibliography{reference} 

@article{CMS:2023pte,
    author = "Hayrapetyan, Aram and others",
    collaboration = "CMS",
    title = "{Search for the lepton-flavor violating decay of the Higgs boson and additional Higgs bosons in the e$\mu$ final state in proton-proton collisions at $\sqrt{s}$ = 13 TeV}",
    eprint = "2305.18106",
    archivePrefix = "arXiv",
    primaryClass = "hep-ex",
    reportNumber = "CMS-HIG-22-002, CERN-EP-2023-061",
    doi = "10.1103/PhysRevD.108.072004",
    journal = "Phys. Rev. D",
    volume = "108",
    number = "7",
    pages = "072004",
    year = "2023"
}

@article{Canonero:2026pul,
    author = "Canonero, Enzo and Cowan, Glen",
    title = "{Discovery Sensitivity for a Counting Experiment with Background Uncertainty}",
    eprint = "2607.29436",
    archivePrefix = "arXiv",
    primaryClass = "physics.data-an",
    month = "7",
    year = "2026"
}

@article{Cowan:2010js,
    author = "Cowan, Glen and Cranmer, Kyle and Gross, Eilam and Vitells, Ofer",
    title = "{Asymptotic formulae for likelihood-based tests of new physics}",
    eprint = "1007.1727",
    archivePrefix = "arXiv",
    primaryClass = "physics.data-an",
    doi = "10.1140/epjc/s10052-011-1554-0",
    journal = "Eur. Phys. J. C",
    volume = "71",
    pages = "1554",
    year = "2011",
    note = "[Erratum: Eur.Phys.J.C 73, 2501 (2013)]"
}

@article{Ball:2012cx,
    author = "Ball, Richard D. and others",
    title = "{Parton distributions with LHC data}",
    eprint = "1207.1303",
    archivePrefix = "arXiv",
    primaryClass = "hep-ph",
    reportNumber = "EDINBURGH-2012-08, IFUM-FT-997, FR-PHENO-2012-014, RWTH-TTK-12-25, CERN-PH-TH-2012-037, SFB-CPP-12-47",
    doi = "10.1016/j.nuclphysb.2012.10.003",
    journal = "Nucl. Phys. B",
    volume = "867",
    pages = "244--289",
    year = "2013"
}

@article{Koivunen:2023led,
    author = "Koivunen, Niko and Raidal, Martti",
    title = "{Production and decays of 146 GeV flavons into e$\mu$ final state at the LHC}",
    eprint = "2305.00014",
    archivePrefix = "arXiv",
    primaryClass = "hep-ph",
    doi = "10.1007/JHEP11(2023)014",
    journal = "JHEP",
    volume = "11",
    pages = "014",
    year = "2023"
}

@article{Crivellin:2014cta,
    author = "Crivellin, Andreas and Hoferichter, Martin and Procura, Massimiliano",
    title = "{Improved predictions for $\mu\to e$ conversion in nuclei and Higgs-induced lepton flavor violation}",
    eprint = "1404.7134",
    archivePrefix = "arXiv",
    primaryClass = "hep-ph",
    reportNumber = "CERN-PH-TH-2014-068",
    doi = "10.1103/PhysRevD.89.093024",
    journal = "Phys. Rev. D",
    volume = "89",
    pages = "093024",
    year = "2014"
}

@article{Crivellin:2013wna,
    author = "Crivellin, Andreas and Kokulu, Ahmet and Greub, Christoph",
    title = "{Flavor-phenomenology of two-Higgs-doublet models with generic Yukawa structure}",
    eprint = "1303.5877",
    archivePrefix = "arXiv",
    primaryClass = "hep-ph",
    doi = "10.1103/PhysRevD.87.094031",
    journal = "Phys. Rev. D",
    volume = "87",
    number = "9",
    pages = "094031",
    year = "2013"
}

@article{Afik:2023vyl,
    author = "Afik, Yoav and Bhupal Dev, P. S. and Thapa, Anil",
    title = "{Hints of a new leptophilic Higgs sector?}",
    eprint = "2305.19314",
    archivePrefix = "arXiv",
    primaryClass = "hep-ph",
    doi = "10.1103/PhysRevD.109.015003",
    journal = "Phys. Rev. D",
    volume = "109",
    number = "1",
    pages = "015003",
    year = "2024"
}

@article{Alloul:2013bka,
    author = "Alloul, Adam and Christensen, Neil D. and Degrande, C\'eline and Duhr, Claude and Fuks, Benjamin",
    title = "{FeynRules  2.0 - A complete toolbox for tree-level phenomenology}",
    eprint = "1310.1921",
    archivePrefix = "arXiv",
    primaryClass = "hep-ph",
    reportNumber = "CERN-PH-TH-2013-239, MCNET-13-14, IPPP-13-71, DCPT-13-142, PITT-PACC-1308",
    doi = "10.1016/j.cpc.2014.04.012",
    journal = "Comput. Phys. Commun.",
    volume = "185",
    pages = "2250--2300",
    year = "2014"
}

@article{Alwall:2014hca,
    author = "Alwall, J. and Frederix, R. and Frixione, S. and Hirschi, V. and Maltoni, F. and Mattelaer, O. and Shao, H. -S. and Stelzer, T. and Torrielli, P. and Zaro, M.",
    title = "{The automated computation of tree-level and next-to-leading order differential cross sections, and their matching to parton shower simulations}",
    eprint = "1405.0301",
    archivePrefix = "arXiv",
    primaryClass = "hep-ph",
    reportNumber = "CERN-PH-TH-2014-064, CP3-14-18, LPN14-066, MCNET-14-09, ZU-TH-14-14",
    doi = "10.1007/JHEP07(2014)079",
    journal = "JHEP",
    volume = "07",
    pages = "079",
    year = "2014"
}

@article{Bierlich:2022pfr,
    author = "Bierlich, Christian and others",
    title = "{A comprehensive guide to the physics and usage of PYTHIA 8.3}",
    eprint = "2203.11601",
    archivePrefix = "arXiv",
    primaryClass = "hep-ph",
    reportNumber = "LU-TP 22-16, MCNET-22-04, FERMILAB-PUB-22-227-SCD",
    doi = "10.21468/SciPostPhysCodeb.8",
    journal = "SciPost Phys. Codeb.",
    volume = "2022",
    pages = "8",
    year = "2022"
}

@article{MEGII:2025gzr,
    author = "Afanaciev, K. and others",
    collaboration = "MEG II",
    title = "{New limit on the ${\mu ^+ \rightarrow e^+ \gamma }$ decay with the MEG II experiment}",
    eprint = "2504.15711",
    archivePrefix = "arXiv",
    primaryClass = "hep-ex",
    doi = "10.1140/epjc/s10052-025-14906-3",
    journal = "Eur. Phys. J. C",
    volume = "85",
    number = "10",
    pages = "1177",
    year = "2025",
    note = "[Erratum: Eur.Phys.J.C 85, 1317 (2025)]"
}

@article{Altmannshofer:2025nsl,
    author = "Altmannshofer, Wolfgang and Assi, Beno{\^\i}t and Brod, Joachim and Hamer, Nick and Julio, J. and Uttayarat, Patipan and Volkov, Daniil",
    title = "{Electron EDM and {\ensuremath{\Gamma}}({\ensuremath{\mu}} {\textrightarrow} e{\ensuremath{\gamma}}) in the 2HDM}",
    eprint = "2410.17313",
    archivePrefix = "arXiv",
    primaryClass = "hep-ph",
    doi = "10.1007/JHEP06(2025)156",
    journal = "JHEP",
    volume = "06",
    pages = "156",
    year = "2025"
}

@article{Ginzburg:2003fe,
    author = "Ginzburg, I. F. and Ivanov, I. P.",
    title = "{Tree level unitarity constraints in the 2HDM with CP violation}",
    eprint = "hep-ph/0312374",
    archivePrefix = "arXiv",
    month = "12",
    year = "2003"
}

@article{Buonocore:2020nai,
    author = "Buonocore, Luca and Nason, Paolo and Tramontano, Francesco and Zanderighi, Giulia",
    title = "{Leptons in the proton}",
    eprint = "2005.06477",
    archivePrefix = "arXiv",
    primaryClass = "hep-ph",
    doi = "10.1007/JHEP08(2020)019",
    journal = "JHEP",
    volume = "08",
    number = "08",
    pages = "019",
    year = "2020"
}

@article{Gao:2026qnt,
    author = "Gao, Christina and Li, Lingfeng and Xiong, Z. J.",
    title = "{Probing a 146 GeV cLFV scalar using the LHC and low-energy experiments}",
    eprint = "2607.03249",
    archivePrefix = "arXiv",
    primaryClass = "hep-ph",
    month = "7",
    year = "2026"
}

@article{Grimus:2008nb,
    author = "Grimus, W. and Lavoura, L. and Ogreid, O. M. and Osland, P.",
    title = "{The Oblique parameters in multi-Higgs-doublet models}",
    eprint = "0802.4353",
    archivePrefix = "arXiv",
    primaryClass = "hep-ph",
    reportNumber = "UWTHPH-2008-4",
    doi = "10.1016/j.nuclphysb.2008.04.019",
    journal = "Nucl. Phys. B",
    volume = "801",
    pages = "81--96",
    year = "2008"
}

@article{Primulando:2022vip,
    author = "Primulando, R. and Julio, J. and Uttayarat, P.",
    title = "{Minimal Zee model for lepton g-2 and W-mass shifts}",
    eprint = "2211.16021",
    archivePrefix = "arXiv",
    primaryClass = "hep-ph",
    doi = "10.1103/PhysRevD.107.055034",
    journal = "Phys. Rev. D",
    volume = "107",
    number = "5",
    pages = "055034",
    year = "2023"
}

@article{Uttayarat:2025qjl,
    author = "Uttayarat, P. and Julio, J. and Primulando, R.",
    title = "{Novel probes for electron-muon flavor violation from exotic Higgs decays}",
    eprint = "2508.20932",
    archivePrefix = "arXiv",
    primaryClass = "hep-ph",
    doi = "10.1007/JHEP03(2026)260",
    journal = "JHEP",
    volume = "03",
    pages = "260",
    year = "2026"
}

@article{Ivanov:2006yq,
    author = "Ivanov, I. P.",
    title = "{Minkowski space structure of the Higgs potential in 2HDM}",
    eprint = "hep-ph/0609018",
    archivePrefix = "arXiv",
    doi = "10.1103/PhysRevD.75.035001",
    journal = "Phys. Rev. D",
    volume = "75",
    pages = "035001",
    year = "2007",
    note = "[Erratum: Phys.Rev.D 76, 039902 (2007)]"
}

@article{ATLAS:2022vkf,
    author = "Aad, Georges and others",
    collaboration = "ATLAS",
    title = "{A detailed map of Higgs boson interactions by the ATLAS experiment ten years after the discovery}",
    eprint = "2207.00092",
    archivePrefix = "arXiv",
    primaryClass = "hep-ex",
    reportNumber = "CERN-EP-2022-057",
    doi = "10.1038/s41586-022-04893-w",
    journal = "Nature",
    volume = "607",
    number = "7917",
    pages = "52--59",
    year = "2022",
    note = "[Erratum: Nature 612, E24 (2022)]"
}

@article{ATLAS:2016neq,
    author = "Aad, Georges and others",
    collaboration = "ATLAS, CMS",
    title = "{Measurements of the Higgs boson production and decay rates and constraints on its couplings from a combined ATLAS and CMS analysis of the LHC pp collision data at $ \sqrt{s}=7 $ and 8 TeV}",
    eprint = "1606.02266",
    archivePrefix = "arXiv",
    primaryClass = "hep-ex",
    reportNumber = "CERN-EP-2016-100, ATLAS-HIGG-2015-07, CMS-HIG-15-002",
    doi = "10.1007/JHEP08(2016)045",
    journal = "JHEP",
    volume = "08",
    pages = "045",
    year = "2016"
}

@article{ATLAS:2019old,
    author = "Aad, Georges and others",
    collaboration = "ATLAS",
    title = "{Search for the Higgs boson decays $H \to ee$ and $H \to e\mu$ in $pp$ collisions at $\sqrt{s} = 13$ TeV with the ATLAS detector}",
    eprint = "1909.10235",
    archivePrefix = "arXiv",
    primaryClass = "hep-ex",
    reportNumber = "CERN-EP-2019-184",
    doi = "10.1016/j.physletb.2019.135148",
    journal = "Phys. Lett. B",
    volume = "801",
    pages = "135148",
    year = "2020"
}

@article{ParticleDataGroup:2024cfk,
    author = "Navas, S. and others",
    collaboration = "Particle Data Group",
    title = "{Review of particle physics}",
    doi = "10.1103/PhysRevD.110.030001",
    journal = "Phys. Rev. D",
    volume = "110",
    number = "3",
    pages = "030001",
    year = "2024"
}

@article{Primulando:2023ugc,
    author = "Primulando, R. and Julio, J. and Srimanobhas, N. and Uttayarat, P.",
    title = "{A new Higgs boson with electron-muon flavor-violating couplings}",
    eprint = "2304.13757",
    archivePrefix = "arXiv",
    primaryClass = "hep-ph",
    doi = "10.1016/j.physletb.2023.138129",
    journal = "Phys. Lett. B",
    volume = "845",
    pages = "138129",
    year = "2023"
}

@article{CMS:2022dwd,
    author = "Tumasyan, Armen and others",
    collaboration = "CMS",
    title = "{A portrait of the Higgs boson by the CMS experiment ten years after the discovery}",
    eprint = "2207.00043",
    archivePrefix = "arXiv",
    primaryClass = "hep-ex",
    reportNumber = "CMS-HIG-22-001, CERN-EP-2022-039",
    doi = "10.1038/s41586-022-04892-x",
    journal = "Nature",
    volume = "607",
    number = "7917",
    pages = "60--68",
    year = "2022"
}

@article{CMS:2021cox,
    author = "Tumasyan, Armen and others",
    collaboration = "CMS",
    title = "{Search for electroweak production of charginos and neutralinos in proton-proton collisions at $ \sqrt{s} $ = 13 TeV}",
    eprint = "2106.14246",
    archivePrefix = "arXiv",
    primaryClass = "hep-ex",
    reportNumber = "CMS-SUS-19-012, CERN-EP-2021-097",
    doi = "10.1007/JHEP04(2022)147",
    journal = "JHEP",
    volume = "04",
    pages = "147",
    year = "2022"
}

@article{deFavereau:2013fsa,
    author = "de Favereau, J. and Delaere, C. and Demin, P. and Giammanco, A. and Lema\^\i{}tre, V. and Mertens, A. and Selvaggi, M.",
    collaboration = "DELPHES 3",
    title = "{DELPHES 3, A modular framework for fast simulation of a generic collider experiment}",
    eprint = "1307.6346",
    archivePrefix = "arXiv",
    primaryClass = "hep-ex",
    doi = "10.1007/JHEP02(2014)057",
    journal = "JHEP",
    volume = "02",
    pages = "057",
    year = "2014"
}

@article{CMS:2022goy,
    author = "Tumasyan, Armen and others",
    collaboration = "CMS",
    title = "{Searches for additional Higgs bosons and for vector leptoquarks in $\tau\tau$ final states in proton-proton collisions at $\sqrt{s}$ = 13 TeV}",
    eprint = "2208.02717",
    archivePrefix = "arXiv",
    primaryClass = "hep-ex",
    reportNumber = "CMS-HIG-21-001, CERN-EP-2022-137",
    doi = "10.1007/JHEP07(2023)073",
    journal = "JHEP",
    volume = "07",
    pages = "073",
    year = "2023"
}

@article{CMS:2018amk,
    author = "Sirunyan, Albert M and others",
    collaboration = "CMS",
    title = "{Search for a new scalar resonance decaying to a pair of Z bosons in proton-proton collisions at $\sqrt{s}=13 $ TeV}",
    eprint = "1804.01939",
    archivePrefix = "arXiv",
    primaryClass = "hep-ex",
    reportNumber = "CMS-HIG-17-012, CERN-EP-2018-009",
    doi = "10.1007/JHEP06(2018)127",
    journal = "JHEP",
    volume = "06",
    pages = "127",
    year = "2018",
    note = "[Erratum: JHEP 03, 128 (2019)]"
}

@article{Georgi:1978ri,
    author = "Georgi, Howard and Nanopoulos, Dimitri V.",
    title = "{Suppression of Flavor Changing Effects From Neutral Spinless Meson Exchange in Gauge Theories}",
    reportNumber = "HUTP-78/A055",
    doi = "10.1016/0370-2693(79)90433-7",
    journal = "Phys. Lett. B",
    volume = "82",
    pages = "95--96",
    year = "1979"
}

@article{Fukuyama:2021iyw,
    author = "Fukuyama, Takeshi and Mimura, Yukihiro and Uesaka, Yuichi",
    title = "{Models of the muonium to antimuonium transition}",
    eprint = "2108.10736",
    archivePrefix = "arXiv",
    primaryClass = "hep-ph",
    doi = "10.1103/PhysRevD.105.015026",
    journal = "Phys. Rev. D",
    volume = "105",
    number = "1",
    pages = "015026",
    year = "2022"
}

@article{Conlin:2020veq,
    author = "Conlin, Renae and Petrov, Alexey A.",
    title = "{Muonium-antimuonium oscillations in effective field theory}",
    eprint = "2005.10276",
    archivePrefix = "arXiv",
    primaryClass = "hep-ph",
    reportNumber = "WSU-HEP-2002",
    doi = "10.1103/PhysRevD.102.095001",
    journal = "Phys. Rev. D",
    volume = "102",
    number = "9",
    pages = "095001",
    year = "2020"
}

@article{Hou:1995dg,
    author = "Hou, Wei-Shu and Wong, Gwo-Guang",
    title = "{mu+ e- {\ensuremath{<}}---{\ensuremath{>}} mu- e+ transitions via neutral scalar bosons}",
    eprint = "hep-ph/9504311",
    archivePrefix = "arXiv",
    reportNumber = "NTUTH-95-03",
    doi = "10.1103/PhysRevD.53.1537",
    journal = "Phys. Rev. D",
    volume = "53",
    pages = "1537--1541",
    year = "1996"
}

@article{Willmann:1998gd,
    author = "Willmann, L. and others",
    title = "{New bounds from searching for muonium to anti-muonium conversion}",
    eprint = "hep-ex/9807011",
    archivePrefix = "arXiv",
    reportNumber = "UHD-PI-MY-9812",
    doi = "10.1103/PhysRevLett.82.49",
    journal = "Phys. Rev. Lett.",
    volume = "82",
    pages = "49--52",
    year = "1999"
}

@article{OPAL:2003kcu,
    author = "Abbiendi, G. and others",
    collaboration = "OPAL",
    title = "{Tests of the standard model and constraints on new physics from measurements of fermion pair production at 189-GeV to 209-GeV at LEP}",
    eprint = "hep-ex/0309053",
    archivePrefix = "arXiv",
    reportNumber = "CERN-EP-2003-053",
    doi = "10.1140/epjc/s2004-01595-9",
    journal = "Eur. Phys. J. C",
    volume = "33",
    pages = "173--212",
    year = "2004"
}

@article{CMS:2018eqb,
    author = "Sirunyan, Albert M. and others",
    collaboration = "CMS",
    title = "{Search for supersymmetric partners of electrons and muons in proton-proton collisions at $\sqrt{s}=$ 13 TeV}",
    eprint = "1806.05264",
    archivePrefix = "arXiv",
    primaryClass = "hep-ex",
    reportNumber = "CMS-SUS-17-009, CERN-EP-2018-132",
    doi = "10.1016/j.physletb.2019.01.005",
    journal = "Phys. Lett. B",
    volume = "790",
    pages = "140--166",
    year = "2019"
}

@article{ATLAS:2019lff,
    author = "Aad, Georges and others",
    collaboration = "ATLAS",
    title = "{Search for electroweak production of charginos and sleptons decaying into final states with two leptons and missing transverse momentum in $\sqrt{s}=13$ TeV $pp$ collisions using the ATLAS detector}",
    eprint = "1908.08215",
    archivePrefix = "arXiv",
    primaryClass = "hep-ex",
    reportNumber = "CERN-EP-2019-106",
    doi = "10.1140/epjc/s10052-019-7594-6",
    journal = "Eur. Phys. J. C",
    volume = "80",
    number = "2",
    pages = "123",
    year = "2020"
}
\bibliographystyle{JHEP}
\end{document}